\documentclass[%
 reprint,
 superscriptaddress,
 amsmath,amssymb,
 aps,prl
]{revtex4-1}

\usepackage[dvipdfmx]{graphicx}%
\usepackage{dcolumn}%
\usepackage{bm}%
\usepackage{color}%

\begin{document}

\preprint{APS/123-QED}

\title{Extremely large magnetoresistance induced by hidden three-dimensional Dirac bands in nonmagnetic semimetal InBi}%

\author{K. Okawa}
\email{okawa.k@aist.go.jp}
\affiliation{%
 Laboratory for Materials and Structures, Tokyo Institute of Technology, Kanagawa 226-8503, Japan
}%
\affiliation{National Institute of Advanced Industrial Science and Technology (AIST), National Metrology Institute of Japan (NMIJ), Tsukuba, Ibaraki, Japan 305-8563}
\author{M. Kanou}%
\affiliation{%
 Laboratory for Materials and Structures, Tokyo Institute of Technology, Kanagawa 226-8503, Japan
}%
\author{H. Namiki}%
\affiliation{%
 Laboratory for Materials and Structures, Tokyo Institute of Technology, Kanagawa 226-8503, Japan
}%
\author{T. Sasagawa}%
\email{sasagawa.t.aa@m.titech.ac.jp}
\affiliation{%
 Laboratory for Materials and Structures, Tokyo Institute of Technology, Kanagawa 226-8503, Japan
}%

\date{\today}

\begin{abstract}
Extremely large positive magnetoresistance (XMR) was found in a nonmagnetic semimetal InBi. Using several single crystals with different residual resistivity ratios ({\ensuremath{RRR}}s), we revealed that the XMR strongly depended on the {\ensuremath{RRR}} (sample quality). Assuming that there were no changes in effective mass $m^*$ and carrier concentrations in these single crystals, this dependence was explained by a semiclassical two-carrier model. First-principle calculations including the spin-orbit interactions (SOI) unveiled that InBi had a compensated carrier balance and SOI-induced ``hidden'' three-dimensional (3D) Dirac bands at the M and R points. Because the small $m^*$ and the large carrier mobilities will be realized, these hidden 3D Dirac bands should play an important role for the XMR in InBi. We suggest that this feature can be employed as a novel strategy for the creation of XMR semimetals. 

\end{abstract}

\maketitle


\section{\label{sec:level1}Introduction}
Electrical-resistance changes in solids upon application of external magnetic fields are referred to as magnetoresistance (MR) effects (MR $\equiv \Delta\rho = (\rho(H) - \rho(0)) / \rho(0)$). In particular, magnetic compounds, such as Fe/Cr multilayers and Mn-based oxides, exhibit a giant MR (GMR, $\sim 10^2\%$) \cite{Binasch1,Baibich2} and colossal MR (CMR, $\sim 10^4\%$) \cite{Morimoto3} originated from the suppression of spin disorder scattering (thus negative MR). The MR effects have contributed to the development of spintronics and its application. On the other hand, in the case of nonmagnetic compounds, positive MR is induced by an increased carrier scattering caused by the Lorenz force in a magnetic field. Conventional metals exhibit small MRs (on the order of a few percentages) owing to the weak-scattering. A large MR was found in semi-metallic bismuth and graphite, which have been investigated in the past few decades \cite{Kapitza4,Alers5,Mcclure6,Du7}. 

Recently, several nonmagnetic (semi)metals have been reported with extremely large MRs (XMR: $10^5-10^8 \%$), following the report on XMR in WTe$_2$, which has a good balance of hole and electron carrier concentrations \cite{Ali8} and was later turned out to be a topological Weyl semimetal \cite{Soluyanov9}. The good balance of carrier concentrations, referred to as ``compensated'' carriers, is considered as a common feature of XMR materials. From the semiclassical two-band model, the MR in the compensated semimetals is expressed as MR $\sim \mu_e \mu_h H^2$, where $\mu_e$ and $\mu_h$ are the mobilities of the electron and hole carriers, respectively.The important feature in this model is that the MR is expected to show non-saturating behavior at high magnetic fields and can reach extremely high values, in contrast to the GMR and CMR in typical magnetic compounds. Therefore, XMR materials have high carrier mobilities. Three-dimensional (3D) Dirac semimetals and Weyl semimetals, which have high mobilities and small effective masses owing to the linear (Dirac or Weyl) band dispersions near the Fermi level, have also exhibited XMR. 3D Dirac semimetals (Cd$_3$As$_2$ \cite{Liang10} and ZrSiS \cite{Singha11}), type-I Weyl semimetals ($XPn$ ($X$ = Ta, Nb; $Pn$ = P, As) \cite{Zhang12,Ghimire13,Zhang14,Shekhar15}), and type-II Weyl semimetals (WTe$_2$ \cite{Ali8} and MoTe$_2$ \cite{Rhodes16}) have been reported to exhibit high mobilities and XMR. These materials possess strong spin-orbit interactions (SOIs), realizing relativistic electronic structures (Dirac/Weyl dispersions).

\begin{figure}[b]
\centering
\includegraphics[width=8.5cm,clip]{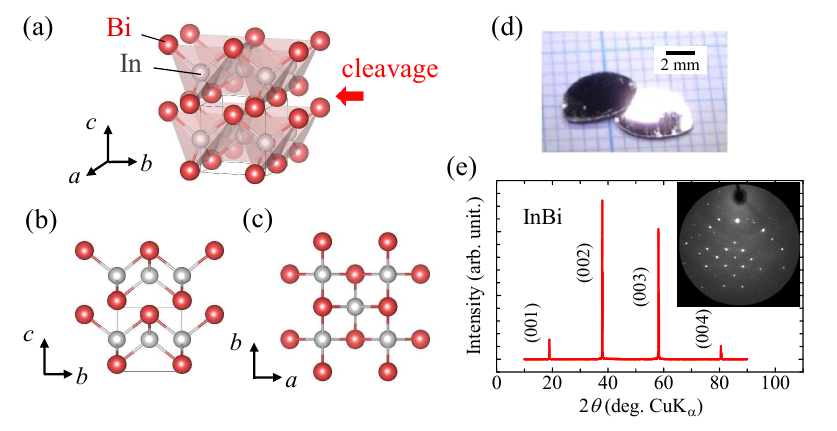}
\caption{\label{fig:wide}(a) Crystal structure of the layered semimetal InBi. InBi has a cleavage plane between InBi$_4$-tetrahedron layers. Perspective views of the (b) $bc$-plane and (c) $ab$-plane. (d) Grown single crystals of InBi. (e) X-ray diffraction pattern from the cleaved surface of the single crystal. The inset shows the two-dimensional pattern obtained from the X-ray diffraction of the single crystal.}
\label{fig1}
\end{figure}

\begin{figure*}[t]
\centering
\includegraphics[width=17cm,clip]{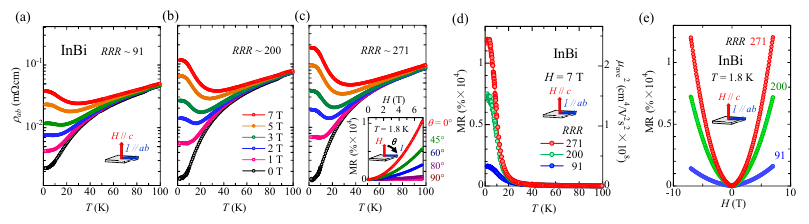}
\caption{\label{fig:wide} (a-c) Temperature dependences of the $ab$-plane resistivity $\rho_{ab}$ under various magnetic fields $H (0\sim7$ T) in the InBi single crystals with $RRR$s of = 91, 200, and 271. The inset shows the magnetic-field dependence of the MR for tilt angles $\theta$ from $0^{\circ}$ to $90^{\circ}$ for a large-$RRR$ crystal ($RRR \sim 271$) at $T = 1.8$ K. (d) Temperature dependence of the MR and the corresponding average carrier mobility $\mu^2_{\rm{ave}}$ for the crystals with different $RRR$s at $H = 7$ T, where the applied field is parallel to the $c$-axis.
(e) Magnetic-field dependence of the MR for the crystals with different $RRR$s at $T = 1.8$ K.}
\label{fig2}
\end{figure*}

In this study, we investigated semimetal InBi, which is nonmagnetic (diamagnetic) down to $T \sim 2$ K \cite{Nishimura17}. InBi has been reported to exhibit superconductivity under hydrostatic pressure ($T_{\rm c} \sim 1.4$ K, $P \sim 0.8$ GPa) \cite{Tissen17}. The semimetallic electronic structure of InBi has been revealed \cite{Schirber18,Maezawa19}, indicating InBi has a good potential to have a desired carrier concentration balance. In addition, a strong atomic SOI of Bi exists in this compound. The two key features for the XMR, a good balance of carrier concentrations and strong SOI, coexist in InBi. Therefore, InBi is a promising candidate for an XMR semimetal. Magneto-transport measurements have been reported, and a large MR ($\sim 3.5 \times 10^3\%$, $T \sim 4.2$ K, $H \sim 10$ T) has been observed. The results of Shubnikov-de-Haas oscillations indicated that InBi is a compensated metal with closed orbits \cite{Maezawa19}. Very recently, exotic topological electric structures (such as a type-II Dirac point and a Dirac nodal line) below the $E_{\rm F}$ in InBi were discovered by angle resolved photoemission spectroscopy measurements \cite{Ekahana20}. The magneto-transport properties of InBi may be related to the topological electric structures. 

In order to investigate in detail the magneto-transport properties of InBi, a good candidate for an XMR material, high-quality single crystals were grown by the modified horizontal Bridgman technique. By performing magneto-transport measurements, the normal and MR properties were evaluated. By performing angular dependent magneto-transport measurements, the anisotropic magneto-transport properties were investigated. In addition, using single crystals with different sample qualities ($RRR$), the relation between the MR and sample quality was revealed. By performing relativistic first-principle calculations, the bulk electronic structures were evaluated, and the origin of the MR in InBi was discussed.

\section{material and method}

InBi is a layered semimetal with a centrosymmetric tetragonal structure (space group $P$4/$nmm$), with cleavage property between the stacked Bi-In-Bi layers (Figs.~\ref{fig1} (a)\,-\,(c)). Single crystals of InBi were grown by the modified horizontal Bridgman technique. In and Bi were mixed in a molar ratio of 1:1. The mixture was sealed in an evacuated quartz tube, and preliminarily reacted by heating until it completely melted. Using the obtained ingot, the crystal growth was performed in a home-built horizontal moving tube furnace. The heating unit was maintained at $200^{\circ}$C (the melting temperature of InBi is $110^{\circ}$C), and moved in the lateral direction of the sealed quartz tube with a speed of 1 mm/h. The obtained single crystals had a good cleavage property along the $ab$-plane, producing flat surfaces with sizes of $\sim 5 \times 5$ ${\rm mm}^2$ (Fig.~\ref{fig1}(d)). The X-ray diffraction pattern from the cleaved surface of the InBi single crystal is shown in Fig.~\ref{fig1}(e). The observed peaks were only from the (00$l$) reflections, indicating that the cleavage plane is the $ab$-plane. This is consistent with the existence of the weak bonding between Bi-In-Bi layers, expected from the crystal structure shown in Fig.~\ref{fig1}(a). The single crystal X-ray diffraction (SXRD) pattern from the cleaved surface of the InBi is shown in the inset of Fig.~\ref{fig1}(c). From the SXRD data, the lattice constants were determined to be $a = b = 4.991 {\rm{\AA}}$, and $c = 4.776 {\rm{\AA}}$. 
For magneto-transport measurements, rectangular bar-shaped crystals ($\sim 1 \times 0.5 \times 0.01 $ ${\rm{mm}}^3$, the shortest along the $c$-axis) were prepared by cleaving and cutting. Resistivity was measured with the standard four-probe method. 

\section{results and discussion}

In order to systematically investigate the relation between the MR and sample quality, Figs.~\ref{fig2}(a)\,-\,(c) showed the temperature dependencies of the $ab$-plane resistivity $\rho_{ab}$ in InBi (the residual resistivity ratio [$RRR \equiv \rho$(300 K)/$\rho$(4 K)] is $\sim$ (a) 91, (b) 200, and (c) 271) under various magnetic fields $H (0\sim7$ T); the current is along the $ab$-plane, while the applied field is parallel to the $c$-axis ($I \parallel ab, H \parallel c$). All of the samples exhibited field-induced metal-insulator-''like'' transitions (the resistivity upturn below 30 K), as similar to other XMR materials \cite{Ali8,Singha11,Zhang14,Shekhar15,Tafti21}. This upturn is caused not by the transition to an insulating phase but by the rapid increase of the MR below 30 K, as shown in Fig.~\ref{fig2}(d). As described later, in semiclassical two-carrier model, the MR is related to the average of the carrier mobility $\mu^2_{\rm{ave}}$. Therefore, the dramatic change in  $\rho(T)$ under the magnetic field at low temperature is caused by the strong temperature dependence of the carrier mobility. Fig.~\ref{fig2}(d) clearly indicates that the MR in InBi below 30 K strongly depends on the $RRR$ (sample quality). The field dependences of the MR for several $RRR$s at 1.8 K are shown in Fig.~\ref{fig2}(e). All of the samples exhibited a quadratic field dependence of the MR (MR $\propto H^2$). The value of the MR at 7 T reached $1.2 \times 10^4\%$ for $RRR \sim 271, 7.2 \times 10^3\%$ for $RRR \sim 200,$ and $1.6 \times 10^3\%$ for $RRR \sim 91$. The field dependence of the MR in InBi ($RRR \sim 271$) at 1.8 K (the current is along the $ab$-plane, while the applied field is at an angle of $\theta = 0^{\circ}, 45^{\circ}, 60^{\circ}, 80^{\circ},$ and $90^{\circ}$ with respect to the $c$-axis) is shown in the inset of Fig.~\ref{fig2}(c). The values of the MR decreased with the increase of the angle from $0^{\circ} (H \perp I$) to $90^{\circ} (H \parallel I)$ $(5.4 \times 10^3\%$ at $45^{\circ}$, $2.5 \times 10^3\%$ at $60^{\circ}$, and $5.3 \times 10^2\%$ at $80^{\circ})$. The minimum MR at $\theta \sim 90^{\circ}$ (which in principle should be zero) was as large as $\sim 50\%$ within the attainable accuracy of the rotator alignment, which in turn indicated the largeness of the MR in InBi.

\begin{figure}[t]
\centering
\includegraphics[width=8.5cm,clip]{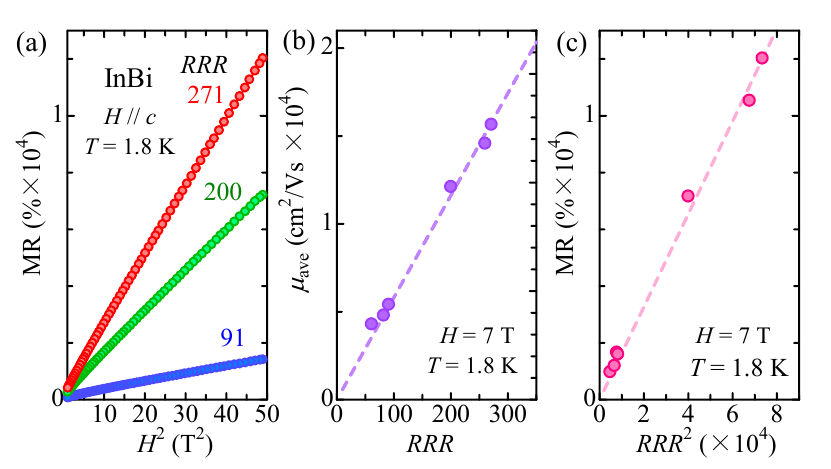}
\caption{\label{fig:wide}(a) MR as a function of $H^2$ for the crystals with different $RRR$s of $\sim$ 91, 200, and 271, at $T = 1.8$ K, where the applied field is parallel to the $c$-axis. (b) $RRR$ dependence of the average carrier mobility $\mu_{\rm{ave}}$ at $T = 1.8$ K and $H = 7$ T. (c) MR as a function of $RRR^2$ at $T = 1.8$ K and $H = 7$ T.}
\label{fig3}
\end{figure}

Figure~\ref{fig3}(a) shows the squared-field ($H^2$) dependence of the MR of the samples with $RRR$s of approximately 91, 200, and 271 at 1.8 K; the current is along the $ab$-plane, while the applied field is parallel to the $c$-axis ($I \parallel ab, H \parallel c$). The MR curves for the different $RRR$s reveal the linear dependence on the squared field $H^2$. These results can be used to semiquantitatively evaluated the carrier mobility. In the framework of the semiclassical two-band model, the electrical conductivity tensor $\sigma$ is described as:
\begin{eqnarray}
\sigma=e\left[\frac{n_{e}\mu_{e}}{1+i\mu_{e}H}+\frac{n_{h}\mu_{h}}{1-i\mu_{h}H}\right],
\label{eq1}
\end{eqnarray}
where $n_e$($n_h$) is the carrier concentration for electrons (holes), and $\mu_e$($\mu_h$) is the carrier mobility for electrons (holes). The transverse MR, $\rho_{xx}$($H$), is: 
\begin{flalign}
&\rho_{xx}(H)=Re(\rho) \nonumber \\
&= \frac{1}{e}\frac{(n_h\mu_h+n_e\mu_e)+(n_h\mu_e+n_e\mu_h)\mu_h\mu_eH^2}{(n_e\mu_h+n_e\mu_e)^2+(n_h-n_e)^2H^2}.
\label{eq2}
\end{flalign}
Assuming a perfect charge compensation ($n_e = n_h$), as in the compensated semimetals, Eq.~\ref{eq2} can be rewritten as:
\begin{eqnarray}
\rho_{xx}(H)=\frac{1}{e}\frac{1+\mu_h\mu_eH^2}{n(\mu_h+\mu_e)}=\rho_{xx}(0)(1+\mu_h\mu_eH^2).
\label{eq3}
\end{eqnarray}
Using this equation, the MR can be expressed as: 
\begin{eqnarray}
MR=\frac{\rho_{xx}(H)-\rho_{xx}(0)}{\rho_{xx}(0)}=\mu_h\mu_eH^2.
\label{eq4}
\end{eqnarray}
This relation indicates that the MR depends on the carrier mobilities and squared field. When the average of the hole and electron mobilities is defined as $\mu_{\rm{ave}}^{2}\equiv \mu_{n}\mu_{e}$, the MR can be simply described as MR $= \mu_{\rm{ave}}^{2}H^2$. According to the relation, $\mu_{\rm{ave}}$ in the samples can be estimated by the slope in Fig.~\ref{fig3}(a). The estimated values of $\mu_{\rm{ave}}$ at 1.8 K are $1.6 \times 10^{4}$ cm$^{2}$/Vs for $RRR \sim 271, 1.2 \times 10^{4}$ cm$^2$/Vs for $RRR \sim 200$, and $5.4 \times 10^{3}$ cm$^{2}$/Vs for $RRR \sim 91$. These values are by far larger than the corresponding values of conventional metals. On the other hand, in the pronounced XMR materials, the values at 2 K and 9 T are significantly larger than those of InBi; for example, for WTe$_2$, $\mu_{\rm{ave}}$ is $1.6 \times 10^5$ cm$^2$/Vs \cite{Ali22}.

The residual resistivity $\rho_{0}$ is related to the relaxation time $\tau : \rho_{0} = 1/n_{e}e\mu_{e} = m^{*}/n_{e}e^2\tau$, where $m^*$ is the effective mass. From this relation, the $RRR$ dependence on $\tau^{-1}$ and $\mu_{\rm{ave}}$, can be obtained, assuming that there are no changes in $m^{*}$ and carrier concentrations in the single crystals with different $RRR$s:
\begin{eqnarray}
RRR \propto \tau^{-1} \propto \mu_{\rm{ave}}
\label{eq5}
\end{eqnarray}

The $RRR$ dependence of $\mu_{\rm{ave}}$ in InBi at 1.8 K under 7 T is shown in Fig.~\ref{fig3}(b). The results indicate that $\mu_{\rm{ave}}$ linearly depends on the $RRR$. Eq.~\ref{eq5} reveals that the MR linearly increases with $\mu_{\rm{ave}}$, and thus the MR exhibits a linear relationship with the squared $RRR$. Figure~\ref{fig3}(c) shows the squared-$RRR$ dependence of the MR, demonstrating the linear-dependence between the MR and squared-$RRR$ in InBi, consistent with the above discussion. This relation indicates that the MR in InBi can be further increased by preparing higher-quality (larger $RRR$) single crystals. Therefore, considering the relatively small $RRR$ in InBi (271 at most), InBi have the potential to show comparable XMR as found in topological Dirac/Weyl semimetals (WTe$_{2}, 1.8 \times 10^{6}\%$ at 2 K, 9 T with $RRR \sim 1,250$ \cite{Ali22}; NbSb$_2, 1.3 \times 10^5\%$ at 2 K, 9 T with $RRR \sim 450$ \cite{Wang23}. 

\begin{figure*}[t]
\centering
\includegraphics[width=10.5cm,clip]{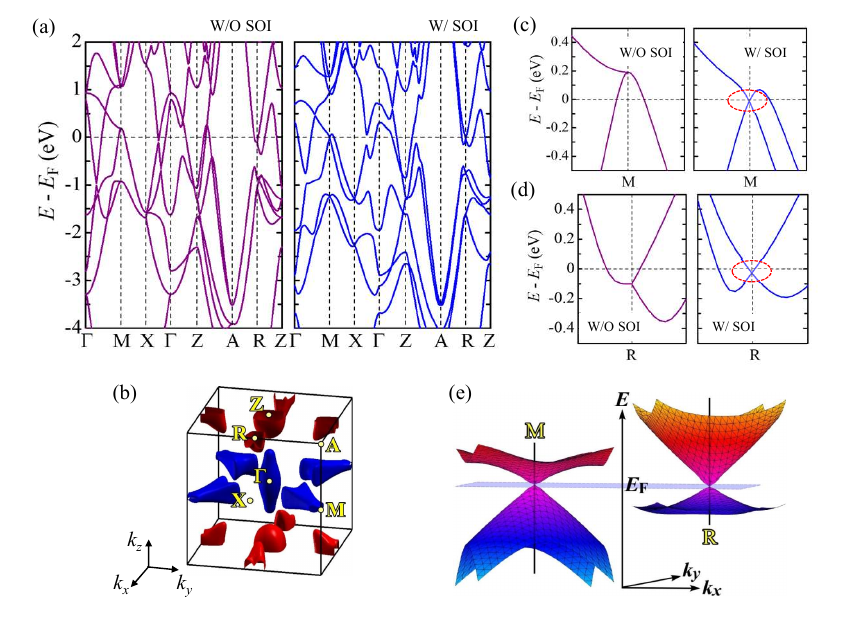}
\caption{\label{fig:wide} (a) Electronic band structures of InBi (left) without SOI and (right) with SOI. (b) 3D Fermi surfaces, with the tetragonal first Brillouin zone, the symmetry points are labeled according to the standard notation. Enlarged views of the electronic band structure near the Fermi level ($E_{\rm{F}}$) around the (c) M and (d) R points (left) without SOI and (right) with SOI. (e) 3D plots of energy versus $xy$-momentum visualizing the hidden 3D Dirac bands around the M and R points with SOI.}
\label{fig4}
\end{figure*}

In order to check possible contributions of topological electronic states near the Fermi level, if any, to XMR in InBi, relativistic first-principle calculations were performed. First-principle calculations within the framework of the density functional theory were performed using the full-potential linearized augmented plane-wave (FP-LAPW) method, as implemented in the WIEN2k code, with the generalized gradient approximation (GGA) and the Perdew-Burke-Ernzerhof (PBE) exchange-correlation functional. The SOI was included as a second variational step with a basis of scalar relativistic eigenfunctions. 

The calculated electronic band structures of InBi are shown in Fig.~\ref{fig4}(a), (left) without and (right) with SOI. InBi exhibited semimetallic band structures. Figure~\ref{fig4}(b) shows the calculated Fermi surfaces with SOI in the first Brillouin zone. The hole pockets around the $\Gamma$ point and zone corners (around the M point) are shown with blue color, while the electron pockets around the Z point and zone corners (around the R point) are shown with red color. The sizes of the hole and electron pockets appeared identical, suggesting that InBi is a compensated semimetal. In addition, the band structure considerably changed in the vicinity of the Fermi energy $E_{\rm{F}}$ upon the inclusion of the SOI. Therefore, it is essential to take into account the relativistic effects in order for the proper understanding of the magneto-transport properties of InBi. 

In particular, we focused on the band dispersions around the high symmetrical M and R points. Figures~\ref{fig4}(c) and~\ref{fig4}(d) show enlarged views of the electronic band structures near $E_{\rm{F}}$ around the (c) M and (d) R points (left) without SOI and (right) with SOI. Without SOI, InBi does not have any band dispersions with a small effective mass $m^*$, related to the high carrier mobility $\mu$. Surprisingly, once SOI was turned on, Dirac-like band dispersions with the crossing point at $E_{\rm{F}}$ were created at both M (hole) and R (electron) points. This is because the orbital degeneracy along the X-M and A-R momentum is resolved by SOI, while it remains at the M and R points. By the 3D plots as shown in Fig.~\ref{fig4}(e), the Dirac-like conical shapes of the energy-momentum relation around the M and R points can be clearly visualized. Therefore, in addition to the compensated carrier balance, the XMR in this compound could be attributed to the small $m^*$ and the large $\mu$ stemming from the SOI-induced ``hidden'' 3D Dirac bands, analogous to the cases of other topological 3D Dirac semimetals such as Cd$_3$As$_2$ and ZrSiS \cite{Liang10,Singha11}. 

\section{summary}

In summary, the magneto-transport properties of a semimetal, InBi, were investigated. The XMR of $\sim 1.2 \times 10^4\%$ at 1.8 K and 7 T was observed in the single crystal with $RRR \sim 271$. The MR exhibited a classical parabolic dependence on the magnetic field without saturation. A strong correlation was observed between the MR and $RRR$. The relation between the MR, $\mu_{\rm{ave}}$, and $RRR$ was explained based on the semiclassical two-band model for compensated semimetals. Based on the experimental results and the plausible model, further drastic (i.e. quadratic) enhancement of the XMR in InBi is anticipated by increasing the $RRR$. First-principle calculations were performed to explain the XMR in InBi. The electronic band structure in InBi exhibited semimetallic features. The Fermi surfaces were composed of the two electron and two hole pockets with similar sizes, indicating that InBi is a compensated semimetal. Furthermore, the 3D Dirac-like bands crossing the Fermi level was found to emerge at the M and R points due to the strong SOI in this material. Therefore, the XMR in InBi was attributed to the large mobilities (small effective mass) of the SOI-induced ``hidden'' 3D Dirac-bands as well as the good balance of the electrons and holes. By considering the lattice symmetry and the orbital degeneracy, it should be possible to search and design the SOI-induced 3D Dirac states, thereby this would serve as a novel strategy for the exploration of XMR semimetals.

\section{acknowledgments}
This work was supported by a CREST project (Grant No. JPMJCR16F2) from Japan Science and Technology Agency (JST) and a Grants-in-Aid for Scientific Research (B) (Grant No. 16H03847) from Japan Society for the Promotion of Science (JSPS). K.O. and M.K. were supported by a research fellowship for young scientists from JSPS (Grants No. 15J11876 and No. a4J10151).

\nocite{*}

\bibliography{apssamp}%

\begin{thebibliography}{26}
\expandafter\ifx\csname natexlab\endcsname\relax\def\natexlab#1{#1}\fi
\expandafter\ifx\csname bibnamefont\endcsname\relax
  \def\bibnamefont#1{#1}\fi
\expandafter\ifx\csname bibfnamefont\endcsname\relax
  \def\bibfnamefont#1{#1}\fi
\expandafter\ifx\csname citenamefont\endcsname\relax
  \def\citenamefont#1{#1}\fi
\expandafter\ifx\csname url\endcsname\relax
  \def\url#1{\texttt{#1}}\fi
\expandafter\ifx\csname urlprefix\endcsname\relax\def\urlprefix{URL }\fi
\providecommand{\bibinfo}[2]{#2}
\providecommand{\eprint}[2][]{\url{#2}}

\bibitem[{\citenamefont{Binasch}(1989)}]{Binasch1}
\bibinfo{author}{\bibfnamefont{G.}~\bibnamefont{Binasch}},
  \bibinfo{author}{\bibfnamefont{P.}~\bibnamefont{Grunberg}},
  \bibinfo{author}{\bibfnamefont{F.}~\bibnamefont{Saurenbach}}, \bibnamefont{and}
  \bibinfo{author}{\bibfnamefont{W.}~\bibnamefont{Zinn}},
  \bibinfo{journal}{Phys.\ Rev.\ B} \textbf{\bibinfo{volume}{39}},
  \bibinfo{pages}{4828} (\bibinfo{year}{1989}).

\bibitem[{\citenamefont{Baibich}(1988)}]{Baibich2}
\bibinfo{author}{\bibfnamefont{M.}~\bibfnamefont{N.}~\bibnamefont{Baibich}},
  \bibinfo{author}{\bibfnamefont{J.}~\bibfnamefont{M.}~\bibnamefont{Broto}},
  \bibinfo{author}{\bibfnamefont{A.}~\bibfnamefont{Fert}},
  \bibinfo{author}{\bibfnamefont{F.}~\bibfnamefont{Nguyen}~\bibnamefont{Van}~\bibnamefont{Dau}}, 
  \bibinfo{author}{\bibfnamefont{F.}~\bibnamefont{Petroff}},
  \bibinfo{author}{\bibfnamefont{P.}~\bibnamefont{Etienne}},
  \bibinfo{author}{\bibfnamefont{G.}~\bibnamefont{Creuzet}},
  \bibinfo{author}{\bibfnamefont{A.}~\bibnamefont{Friederich}}, \bibnamefont{and}
  \bibinfo{author}{\bibfnamefont{J.}~\bibnamefont{Chazelas}},
  \bibinfo{journal}{Phys.\ Rev.\ Lett.} \textbf{\bibinfo{volume}{61}},
  \bibinfo{pages}{2472} (\bibinfo{year}{1988}).

\bibitem[{\citenamefont{Morimoto}(1996)}]{Morimoto3}
\bibinfo{author}{\bibfnamefont{Y.}~\bibnamefont{Morimoto}},
  \bibinfo{author}{\bibfnamefont{A.}~\bibnamefont{Asamitsu}},
  \bibinfo{author}{\bibfnamefont{H.}~\bibnamefont{Kuwahara}}, \bibnamefont{and}
  \bibinfo{author}{\bibfnamefont{Y.}~\bibnamefont{Tokura}},
  \bibinfo{journal}{Nature} \textbf{\bibinfo{volume}{380}},
  \bibinfo{pages}{141} (\bibinfo{year}{1996}).

\bibitem[{\citenamefont{Kapitza}(1928)}]{Kapitza4}
\bibinfo{author}{\bibfnamefont{P.}~\bibnamefont{Kapitza}},
\bibinfo{journal}{Proc.\ R.\ Soc.\ A} \textbf{\bibinfo{volume}{119}},
\bibinfo{pages}{369} (\bibinfo{year}{1928}).

\bibitem[{\citenamefont{Alers}(1953)}]{Alers5}
\bibinfo{author}{\bibfnamefont{P.}~\bibfnamefont{B.}~\bibnamefont{Alers}} \bibnamefont{and}
\bibinfo{author}{\bibfnamefont{R.}~\bibfnamefont{T.}~\bibnamefont{Webber}},
\bibinfo{journal}{Phys.\ Rev.} \textbf{\bibinfo{volume}{91}},
\bibinfo{pages}{1060} (\bibinfo{year}{1953}).

\bibitem[{\citenamefont{Mcclure}(1967)}]{Mcclure6}
\bibinfo{author}{\bibfnamefont{J.}~\bibfnamefont{W.}~\bibnamefont{Mcclure}} \bibnamefont{and}
\bibinfo{author}{\bibfnamefont{J.}~\bibnamefont{Spry}},
\bibinfo{journal}{Phys.\ Rev.} \textbf{\bibinfo{volume}{165}},
\bibinfo{pages}{809} (\bibinfo{year}{1968}).

\bibitem[{\citenamefont{Du}(2005)}]{Du7}
\bibinfo{author}{\bibfnamefont{X.}~\bibfnamefont{Du}},
\bibinfo{author}{\bibfnamefont{S.}~\bibfnamefont{W.}~\bibfnamefont{Tsai}},
\bibinfo{author}{\bibfnamefont{D.}~\bibfnamefont{L.}~\bibfnamefont{Maslov}}, \bibfnamefont{and}, 
\bibinfo{author}{\bibfnamefont{A.}~\bibfnamefont{F.}~\bibfnamefont{Hebard}},
\bibinfo{journal}{Phys.\ Rev.\ Lett.} \textbf{\bibinfo{volume}{94}},
\bibinfo{pages}{166601} (\bibinfo{year}{2005}).

\bibitem[{\citenamefont{Ali}(2014)}]{Ali8}
\bibinfo{author}{\bibfnamefont{M.}~\bibfnamefont{N.}~\bibfnamefont{Ali}},
\bibinfo{author}{\bibfnamefont{J.}~\bibfnamefont{Xiong}},
\bibinfo{author}{\bibfnamefont{S.}~\bibfnamefont{Flynn}},
\bibinfo{author}{\bibfnamefont{J.}~\bibfnamefont{Tao}},
\bibinfo{author}{\bibfnamefont{Q.}~\bibfnamefont{D.}~\bibfnamefont{Gibson}},
\bibinfo{author}{\bibfnamefont{L.}~\bibfnamefont{M.}~\bibfnamefont{Schoop}},
\bibinfo{author}{\bibfnamefont{T.}~\bibfnamefont{Liang}},
\bibinfo{author}{\bibfnamefont{N.}~\bibfnamefont{Haldolaarachchige}},
\bibinfo{author}{\bibfnamefont{M.}~\bibfnamefont{Hirschberger}},
\bibinfo{author}{\bibfnamefont{N.}~\bibfnamefont{P.}~\bibfnamefont{Ong}},\bibfnamefont{and}, 
\bibinfo{author}{\bibfnamefont{R.}~\bibfnamefont{J.}~\bibfnamefont{Cava}},
\bibinfo{journal}{Nature} \textbf{\bibinfo{volume}{514}},
\bibinfo{pages}{205} (\bibinfo{year}{2014}).

\bibitem[{\citenamefont{Soluyanov}(2015)}]{Soluyanov9}
\bibinfo{author}{\bibfnamefont{A.}~\bibfnamefont{A.}~\bibfnamefont{Soluyanov}},
\bibinfo{author}{\bibfnamefont{D.}~\bibfnamefont{Gresch}},
\bibinfo{author}{\bibfnamefont{D.}~\bibfnamefont{Wang}},
\bibinfo{author}{\bibfnamefont{Q-S.}~\bibfnamefont{Wu}},
\bibinfo{author}{\bibfnamefont{M.}~\bibfnamefont{Troyer}},
\bibinfo{author}{\bibfnamefont{X.}~\bibfnamefont{Dai}}, \bibfnamefont{and}, 
\bibinfo{author}{\bibfnamefont{B.}~\bibfnamefont{A.},~\bibfnamefont{Bernevig}},
\bibinfo{journal}{Nature} \textbf{\bibinfo{volume}{527}},
\bibinfo{pages}{495} (\bibinfo{year}{2015}).

\bibitem[{\citenamefont{Liang}(2015)}]{Liang10}
\bibinfo{author}{\bibfnamefont{T.}~\bibfnamefont{Liang}},
\bibinfo{author}{\bibfnamefont{Q.}~\bibfnamefont{Gibson}},
\bibinfo{author}{\bibfnamefont{M.}~\bibfnamefont{N.}~\bibfnamefont{Ali}},
\bibinfo{author}{\bibfnamefont{M.}~\bibfnamefont{Liu}},
\bibinfo{author}{\bibfnamefont{R.}~\bibfnamefont{J.}~\bibfnamefont{Cava}}, \bibfnamefont{and}, 
\bibinfo{author}{\bibfnamefont{N.}~\bibfnamefont{P.}~\bibfnamefont{Ong}},
\bibinfo{journal}{Nat.\ Mat.} \textbf{\bibinfo{volume}{14}},
\bibinfo{pages}{280} (\bibinfo{year}{2014}).

\bibitem[{\citenamefont{Singha}(2017)}]{Singha11}
\bibinfo{author}{\bibfnamefont{R.}~\bibfnamefont{Singha}},
\bibinfo{author}{\bibfnamefont{A.}~\bibfnamefont{Pariari}},
\bibinfo{author}{\bibfnamefont{B.}~\bibfnamefont{Satpati}}, \bibfnamefont{and}, 
\bibinfo{author}{\bibfnamefont{P.}~\bibfnamefont{Mandal}},
\bibinfo{journal}{PNAS} \textbf{\bibinfo{volume}{114}},
\bibinfo{pages}{2468} (\bibinfo{year}{2017}).

\bibitem[{\citenamefont{Zhang}(2016)}]{Zhang12}
\bibinfo{author}{\bibfnamefont{C.}~\bibfnamefont{L.}~\bibfnamefont{Zhang}},
\bibinfo{author}{\bibfnamefont{S.}~\bibfnamefont{Y.}~\bibfnamefont{Xu}},
\bibinfo{author}{\bibfnamefont{I.}~\bibfnamefont{Belopolski}},
\bibinfo{author}{\bibfnamefont{Z.}~\bibfnamefont{Yuan}},
\bibinfo{author}{\bibfnamefont{Z.}~\bibfnamefont{Lin}},
\bibinfo{author}{\bibfnamefont{B.}~\bibfnamefont{Tong}},
\bibinfo{author}{\bibfnamefont{G.}~\bibfnamefont{Bian}},
\bibinfo{author}{\bibfnamefont{N.}~\bibfnamefont{Alidoust}},
\bibinfo{author}{\bibfnamefont{C.}~\bibfnamefont{C.}~\bibfnamefont{Lee}},
\bibinfo{author}{\bibfnamefont{S.}~\bibfnamefont{M.}~\bibfnamefont{Huang}},
\bibinfo{author}{\bibfnamefont{T.}~\bibfnamefont{R.}~\bibfnamefont{Chang}},
\bibinfo{author}{\bibfnamefont{G.}~\bibfnamefont{Chang}},
\bibinfo{author}{\bibfnamefont{C.}~\bibfnamefont{H.}~\bibfnamefont{Hsu}},
\bibinfo{author}{\bibfnamefont{H.}~\bibfnamefont{T.}~\bibfnamefont{Jeng}},
\bibinfo{author}{\bibfnamefont{M.}~\bibfnamefont{Neupane}},
\bibinfo{author}{\bibfnamefont{D.}~\bibfnamefont{S.}~\bibfnamefont{Sanchez}},
\bibinfo{author}{\bibfnamefont{H.}~\bibfnamefont{Zheng}},
\bibinfo{author}{\bibfnamefont{J.}~\bibfnamefont{Wang}},
\bibinfo{author}{\bibfnamefont{H.}~\bibfnamefont{Lin}},
\bibinfo{author}{\bibfnamefont{C.}~\bibfnamefont{Zhang}},
\bibinfo{author}{\bibfnamefont{H.}~\bibfnamefont{Z.}~\bibfnamefont{Lu}},
\bibinfo{author}{\bibfnamefont{S.}~\bibfnamefont{Q.}~\bibfnamefont{Shen}},
\bibinfo{author}{\bibfnamefont{M.}~\bibfnamefont{Neupert}},
\bibinfo{author}{\bibfnamefont{M.}~\bibfnamefont{Z.}~\bibfnamefont{Hasan}}, \bibfnamefont{and}, 
\bibinfo{author}{\bibfnamefont{S.}~\bibfnamefont{Jia}},
\bibinfo{journal}{Nat.\ Commun.} \textbf{\bibinfo{volume}{7}},
\bibinfo{pages}{10735} (\bibinfo{year}{2016}).

\bibitem[{\citenamefont{Ghimire}(2015)}]{Ghimire13}
\bibinfo{author}{\bibfnamefont{N.}~\bibfnamefont{J.}~\bibfnamefont{Liang}},
\bibinfo{author}{\bibfnamefont{Y.}~\bibfnamefont{Luo}},
\bibinfo{author}{\bibfnamefont{M.}~\bibfnamefont{Neupane}},
\bibinfo{author}{\bibfnamefont{M.}~\bibfnamefont{Liu}},
\bibinfo{author}{\bibfnamefont{D.}~\bibfnamefont{J.}~\bibfnamefont{Williams}},
\bibinfo{author}{\bibfnamefont{E.}~\bibfnamefont{D.}~\bibfnamefont{Bauer}}, \bibfnamefont{and}, 
\bibinfo{author}{\bibfnamefont{F.}~\bibfnamefont{Ronning}},
\bibinfo{journal}{J.\ Phys.:\ Condens.\ Matter} \textbf{\bibinfo{volume}{27}},
\bibinfo{pages}{152201} (\bibinfo{year}{2015}).

\bibitem[{\citenamefont{Zhang}(2015)}]{Zhang14}
\bibinfo{author}{\bibfnamefont{C.}~\bibfnamefont{Zhang}},
\bibinfo{author}{\bibfnamefont{C.}~\bibfnamefont{Guo}},
\bibinfo{author}{\bibfnamefont{H.}~\bibfnamefont{Lu}},
\bibinfo{author}{\bibfnamefont{X.}~\bibfnamefont{Zhang}},
\bibinfo{author}{\bibfnamefont{Z.}~\bibfnamefont{Yuan}},
\bibinfo{author}{\bibfnamefont{Z.}~\bibfnamefont{Lin}},
\bibinfo{author}{\bibfnamefont{J.}~\bibfnamefont{Wang}}, \bibfnamefont{and}, 
\bibinfo{author}{\bibfnamefont{S.}~\bibfnamefont{Jia}},
\bibinfo{journal}{Phys.\ Rev.\ B} \textbf{\bibinfo{volume}{92}},
\bibinfo{pages}{041203(R)} (\bibinfo{year}{2015}).

\bibitem[{\citenamefont{Shekhar}(2015)}]{Shekhar15}
\bibinfo{author}{\bibfnamefont{C.}~\bibfnamefont{Shekhar}},
\bibinfo{author}{\bibfnamefont{A.}~\bibfnamefont{K.}~\bibfnamefont{Nayak}},
\bibinfo{author}{\bibfnamefont{Y.}~\bibfnamefont{Sun}},
\bibinfo{author}{\bibfnamefont{M.}~\bibfnamefont{Schmidt}},
\bibinfo{author}{\bibfnamefont{M.}~\bibfnamefont{Nicklas}},
\bibinfo{author}{\bibfnamefont{I.}~\bibfnamefont{Leermakers}},
\bibinfo{author}{\bibfnamefont{U.}~\bibfnamefont{Zeitler}},
\bibinfo{author}{\bibfnamefont{Y.}~\bibfnamefont{Skourski}},
\bibinfo{author}{\bibfnamefont{J.}~\bibfnamefont{Wosnitza}},
\bibinfo{author}{\bibfnamefont{Z.}~\bibfnamefont{Liu}},
\bibinfo{author}{\bibfnamefont{Y.}~\bibfnamefont{Chen}},
\bibinfo{author}{\bibfnamefont{W.}~\bibfnamefont{Schnelle}},
\bibinfo{author}{\bibfnamefont{H.}~\bibfnamefont{Borrmann}},
\bibinfo{author}{\bibfnamefont{Y.}~\bibfnamefont{Grin}},
\bibinfo{author}{\bibfnamefont{C.}~\bibfnamefont{Felser}}, \bibfnamefont{and}, 
\bibinfo{author}{\bibfnamefont{B.}~\bibfnamefont{Yan}},
\bibinfo{journal}{Nat.\ Phys.} \textbf{\bibinfo{volume}{11}},
\bibinfo{pages}{645} (\bibinfo{year}{2015}).

\bibitem[{\citenamefont{Rhodes}(2017)}]{Rhodes16}
\bibinfo{author}{\bibfnamefont{D.}~\bibfnamefont{Rhodes}},
\bibinfo{author}{\bibfnamefont{R.}~\bibfnamefont{Schonemann}},
\bibinfo{author}{\bibfnamefont{N.}~\bibfnamefont{Aryal}},
\bibinfo{author}{\bibfnamefont{Q.}~\bibfnamefont{Zhou}},
\bibinfo{author}{\bibfnamefont{Q.}~\bibfnamefont{R.}~\bibfnamefont{Zhang}},
\bibinfo{author}{\bibfnamefont{E.}~\bibfnamefont{Kampert}},
\bibinfo{author}{\bibfnamefont{Y-C.}~\bibfnamefont{Chiu}},
\bibinfo{author}{\bibfnamefont{Y.}~\bibfnamefont{Lai}},
\bibinfo{author}{\bibfnamefont{Y.}~\bibfnamefont{Shimura}},
\bibinfo{author}{\bibfnamefont{G.}~\bibfnamefont{T.}~\bibfnamefont{McCandless}},
\bibinfo{author}{\bibfnamefont{J.}~\bibfnamefont{Y.}~\bibfnamefont{Chan}},
\bibinfo{author}{\bibfnamefont{D.}~\bibfnamefont{W.}~\bibfnamefont{Paley}},
\bibinfo{author}{\bibfnamefont{J.}~\bibfnamefont{Lee}},
\bibinfo{author}{\bibfnamefont{A.}~\bibfnamefont{D.}~\bibfnamefont{Finke}},
\bibinfo{author}{\bibfnamefont{J.}~\bibfnamefont{P.}~\bibfnamefont{C.}~\bibfnamefont{Ruff}},
\bibinfo{author}{\bibfnamefont{S.}~\bibfnamefont{Das}},
\bibinfo{author}{\bibfnamefont{E.}~\bibfnamefont{Manousakis}}, \bibfnamefont{and}, 
\bibinfo{author}{\bibfnamefont{L.}~\bibfnamefont{Balicas}},
\bibinfo{journal}{Phys.\ Rev.\ B} \textbf{\bibinfo{volume}{96}},
\bibinfo{pages}{165134} (\bibinfo{year}{2017}).

\bibitem[{\citenamefont{Nishimura}(2003)}]{Nishimura17}
\bibinfo{author}{\bibfnamefont{K.}~\bibfnamefont{Nishimura}},
\bibinfo{author}{\bibfnamefont{T.}~\bibfnamefont{Yasukawa}},
\bibfnamefont{and}, 
\bibinfo{author}{\bibfnamefont{K.}~\bibfnamefont{Mori}},
\bibinfo{journal}{Physica\ B} \textbf{\bibinfo{volume}{329}},
\bibinfo{pages}{1399} (\bibinfo{year}{2003}).

\bibitem[{\citenamefont{Tissen}(1998)}]{Tissen17}
\bibinfo{author}{\bibfnamefont{V.}~\bibfnamefont{G.}~\bibfnamefont{Tissen}},
\bibinfo{author}{\bibfnamefont{V.}~\bibfnamefont{F.}~\bibfnamefont{Degtyareva}},
\bibinfo{author}{\bibfnamefont{M.}~\bibfnamefont{V.}~\bibfnamefont{Nefedova}},
\bibinfo{author}{\bibfnamefont{E.}~\bibfnamefont{G.}~\bibfnamefont{Ponyatovski}},
\bibfnamefont{and}, 
\bibinfo{author}{\bibfnamefont{W.}~\bibfnamefont{B.}~\bibfnamefont{Holzapfel}},
\bibinfo{journal}{J.\ Phys.:\ Condens.\ Matter} \textbf{\bibinfo{volume}{10}},
\bibinfo{pages}{7303} (\bibinfo{year}{1998}).

\bibitem[{\citenamefont{Schirber}(1976)}]{Schirber18}
\bibinfo{author}{\bibfnamefont{J.}~\bibfnamefont{E.}~\bibfnamefont{Schirber}}, \bibfnamefont{and}, 
\bibinfo{author}{\bibfnamefont{J.}~\bibfnamefont{P.}~\bibfnamefont{Van}~\bibfnamefont{Dyke}},
\bibinfo{journal}{Phys.\ Rev.\ B} \textbf{\bibinfo{volume}{15}},
\bibinfo{pages}{890} (\bibinfo{year}{1976}).

\bibitem[{\citenamefont{Maezawa}(1982)}]{Maezawa19}
\bibinfo{author}{\bibfnamefont{K.}~\bibfnamefont{Maezawa}},
\bibinfo{author}{\bibfnamefont{Y.}~\bibfnamefont{Saito}},
\bibinfo{author}{\bibfnamefont{T.}~\bibfnamefont{Mizushima}},
\bibfnamefont{and}, 
\bibinfo{author}{\bibfnamefont{S.}~\bibfnamefont{Wakabayashi}},
\bibinfo{journal}{J.\ Phys.\ Soc.\ Jpn.} \textbf{\bibinfo{volume}{51}},
\bibinfo{pages}{3601} (\bibinfo{year}{1982}).

\bibitem[{\citenamefont{Ekahana}(2017)}]{Ekahana20}
\bibinfo{author}{\bibfnamefont{S.}~\bibfnamefont{A.}~\bibfnamefont{Ekahana}},
\bibinfo{author}{\bibfnamefont{S-C.}~\bibfnamefont{Wu}},
\bibinfo{author}{\bibfnamefont{J.}~\bibfnamefont{Jiang}},
\bibinfo{author}{\bibfnamefont{K.}~\bibfnamefont{Okawa}},
\bibinfo{author}{\bibfnamefont{D.}~\bibfnamefont{Prabhakaran}},
\bibinfo{author}{\bibfnamefont{C-C.}~\bibfnamefont{Hwang}},
\bibinfo{author}{\bibfnamefont{S-K.}~\bibfnamefont{Mo}},
\bibinfo{author}{\bibfnamefont{T.}~\bibfnamefont{Sasagawa}},
\bibinfo{author}{\bibfnamefont{C.}~\bibfnamefont{Felser}},
\bibinfo{author}{\bibfnamefont{B.}~\bibfnamefont{Wan}},
\bibinfo{author}{\bibfnamefont{Z.}~\bibfnamefont{Liu}}, \bibfnamefont{and}, 
\bibinfo{author}{\bibfnamefont{Y.}~\bibfnamefont{Chen}},
\bibinfo{journal}{New\ J.\ Phys.} \textbf{\bibinfo{volume}{19}},
\bibinfo{pages}{065007} (\bibinfo{year}{2017}).

\bibitem[{\citenamefont{Tafti}(2016)}]{Tafti21}
  \bibinfo{author}{\bibfnamefont{F.}~\bibfnamefont{F.}~\bibfnamefont{Tafti}},
  \bibinfo{author}{\bibfnamefont{Q.}~\bibfnamefont{Gibson}},
  \bibinfo{author}{\bibfnamefont{S.}~\bibfnamefont{Kushwaha}},
  \bibinfo{author}{\bibfnamefont{J.}~\bibfnamefont{W.}~\bibfnamefont{Krizan}},
  \bibinfo{author}{\bibfnamefont{N.}~\bibfnamefont{Haldolaarachchige}},
  \bibfnamefont{and}, 
  \bibinfo{author}{\bibfnamefont{R.}~\bibfnamefont{J.}~\bibfnamefont{Cava}},
  \bibinfo{journal}{Proc.\ Natl.\ Acad.\ Sci.\ USA} \textbf{\bibinfo{volume}{113}},
  \bibinfo{pages}{E3475} (\bibinfo{year}{2016}).

\bibitem[{\citenamefont{Ali}(2015)}]{Ali22}
  \bibinfo{author}{\bibfnamefont{M.}~\bibfnamefont{N.}~\bibfnamefont{Ali}},
  \bibinfo{author}{\bibfnamefont{L.}~\bibfnamefont{Schoop}},
  \bibinfo{author}{\bibfnamefont{J.}~\bibfnamefont{Xiong}},
  \bibinfo{author}{\bibfnamefont{S.}~\bibfnamefont{Flynn}},
  \bibinfo{author}{\bibfnamefont{Q.}~\bibfnamefont{Gibson}},
  \bibinfo{author}{\bibfnamefont{M.}~\bibfnamefont{Hirschberger}},
  \bibinfo{author}{\bibfnamefont{N.}~\bibfnamefont{P.}~\bibfnamefont{Ong}}, \bibfnamefont{and}, 
  \bibinfo{author}{\bibfnamefont{R.}~\bibfnamefont{J.}~\bibfnamefont{Cava}},
  \bibinfo{journal}{Europhys.\ Lett.} \textbf{\bibinfo{volume}{110}},
  \bibinfo{pages}{67002} (\bibinfo{year}{2015}).

\bibitem[{\citenamefont{Wang}(2014)}]{Wang23}
\bibinfo{author}{\bibfnamefont{K.}~\bibfnamefont{Wang}},
\bibinfo{author}{\bibfnamefont{D.}~\bibfnamefont{Graf}},
\bibinfo{author}{\bibfnamefont{L.}~\bibfnamefont{Li}},
\bibinfo{author}{\bibfnamefont{L.}~\bibfnamefont{Wang}},
\bibfnamefont{and}, 
\bibinfo{author}{\bibfnamefont{C.}~\bibfnamefont{Petrovic}},
\bibinfo{journal}{Sci.\ Rep.} \textbf{\bibinfo{volume}{4}},
\bibinfo{pages}{7328} (\bibinfo{year}{2014}).


\end{thebibliography}

\end{document}